\documentclass{vgtc}                          

\graphicspath{{figs/}{figures/}{pictures/}{images/}{./}} 

\usepackage{times}                     

\usepackage{booktabs}                  
\usepackage{lipsum}                    
\usepackage{mwe}                       

\usepackage{mathptmx}                  

\usepackage{siunitx}
\usepackage{enumitem}
\usepackage{cite}

\usepackage[acronym,shortcuts]{glossaries}
\newacronym{HCI}{HCI}{human-computer interaction}
\newacronym{HMD}{HMD}{head-mounted display}
\newacronym{AR}{AR}{augmented reality}
\newacronym{VR}{VR}{virtual reality}
\newacronym{XR}{XR}{extended reality}
\newacronym{FOV}{FOV}{field of view}
\newacronym{SPV}{SPV}{simulated prosthetic vision}
\newacronym{VPU}{VPU}{vision processing unit}
\newacronym{ONNX}{ONNX}{Open Neural Network Exchange}

\usepackage{xcolor}

\onlineid{2747}

\vgtccategory{Research}

\vgtcinsertpkg

\title{Cross-Modal Guidance for Out-of-View Object Search in Simulated Prosthetic Vision}

\author{Adyah Rastogi\thanks{e-mail: adyah@ucsb.edu} %
\and Apurv Varshney\thanks{e-mail: apurv@ucsb.edu} %
\and Tobias H\"{o}llerer\thanks{e-mail: thollerer@ucsb.edu}
\and Michael Beyeler\thanks{Corresponding author. email: mbeyeler@ucsb.edu}}
\affiliation{\scriptsize University of California, Santa Barbara}

\teaser{
  \centering
  \includegraphics[width=\linewidth, alt={Three-panel overview of the experiment. Panel A shows an Argus II retinal prosthesis with a head-mounted camera, glasses, external electronics, and implanted electrode array. Panel B shows the virtual desk-search environment with a small central SPV field of view nested within a wider VR scene and a target outside the SPV region. Panel C shows example SPV views for baseline, haptic, visual, and audio guidance conditions; haptic and audio cues are indicated by controller and speaker icons, and visual guidance adds peripheral light cues.}]{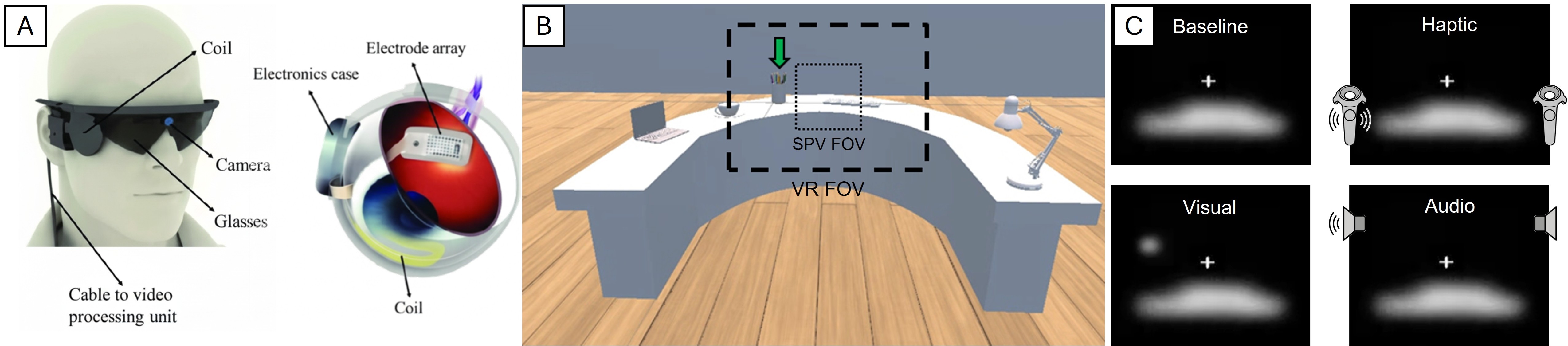}
  \caption{
    Out-of-view object search under simulated prosthetic vision (SPV).
    (A) Retinal prostheses such as Argus~II convert camera input into
    patterned retinal stimulation, producing a sparse visual percept
    (image reused under CC-BY from~\cite{bloch_argus_2019}).
    (B) In our VR task, the guidance system had continuous access to target
    location, while participants viewed the scene through a
    $30^\circ\times30^\circ$ SPV window.
    (C) Visual, haptic, and auditory cues conveyed horizontal target offset
    through laterality and cue rate and were compared with an unguided
    baseline.
    }
  \label{fig:teaser}
}

\abstract{
Out-of-view guidance is well established in virtual and augmented reality, but its effectiveness may depend on the visual bandwidth available to the user.
We test this under simulated prosthetic vision (SPV), where visual guidance must share the same sparse representation used to inspect the scene.
Nineteen participants performed object search under two SPV conditions differing in electrode density and phosphene spread ($10\times10$ and $20\times20$) and four guidance conditions (no guidance, visual, haptic, audio) all driven by the same horizontal target-offset variable.
All three modalities reduced search time and head movement.
The tested auditory and haptic cues produced approximately 25\% faster overall search and 11--13\% faster target acquisition than the visual cue, despite similarly direct orienting trajectories.
The tested haptic and auditory cues also shortened post-acquisition search.
Final head-target angular offset was reduced substantially more in the $10\times10$ SPV condition; there, all three cues also reduced vertical localization error by approximately 45--58\% despite providing no elevation information.
Under severe visual constraints, guidance performance depended on cue implementation and search stage.
}


\begin{document}



\maketitle

\section{Introduction} 

Out-of-view target guidance is a well-established problem in \ac{VR} and
\ac{AR}.
Visual indicators, spatial audio, and vibrotactile cues can all reduce
the cost of finding content outside the current \ac{FOV}
~\cite{baudisch_halo_2003,gustafson_wedge_2008,gruenefeld_comparing_2019,krausComparativeStudyOrientation2020,quinn_augmented_2024}.
These costs become especially important when display \ac{FOV} is
restricted~\cite{ren_evaluating_2016,bork_towards_2018}.

Here we compare visual, auditory, and haptic guidance during search when the visual display itself is severely bandwidth-limited.
Visual neuroprostheses provide an extreme case: implants that seek to restore a form of vision to people who are blind~\cite{fernandez_development_2018,palanker_restoration_2025}.
These devices translate camera input into electrical stimulation,
producing a sparse visual percept with limited spatial resolution and
\ac{FOV}~\cite{palanker_restoration_2025}.
Users must actively scan the environment to bring relevant content into
view~\cite{erickson-davis_what_2021,caspi_combined_2017}, even though the camera may already
capture that content outside the currently represented region
({\bf Figure~\ref{fig:teaser}}).
This mismatch between sensing and perception is particularly relevant for object
search, which is an important unmet need for people who are blind
~\cite{yi_finding_2013,turkstra_assistive_2025}.

Most prosthetic-vision work has focused on improving \emph{what} is
represented once scene content falls within the prosthetic \ac{FOV}; for instance through edge enhancement, depth cues, or semantic
segmentation~\cite{dagnelie_real_2007,sanchez-garcia_semantic_2020,
rasla_relative_2022,han_deep_2021}.
Guidance addresses a complementary problem: getting task-relevant
content into that field in the first place.
Parikh et al.~\cite{parikh_performance_2013} showed that visual saliency
cues can reduce search time and head movement under simulated
prosthetic vision, while Christie et al.~\cite{christie_autonomous_2026}
demonstrated benefits of auditory and haptic feedback for navigation.
Neither study directly compares visual, auditory, and haptic implementations conveying the same underlying guidance variable.

The broader \ac{XR} literature suggests that the choice of modality depends on the demands placed on the visual channel.
Visual guidance can support rapid orienting~\cite{gruenefeld_comparing_2019,marquardt_comparing_2020}, whereas auditory and haptic cues can redirect attention without adding visual content~\cite{biocca_attention_2006,stratmann_exploring_2018}.
Visual augmentation can also introduce clutter or compete with task-relevant information~\cite{gruenefeld_improving_2019,kumaran_impact_2023}.
Whether these tradeoffs hold under prosthetic vision, where guidance and scene content share the same sparse visual representation, is unknown.

We therefore compare visual, auditory, and haptic cues that encode the same horizontal target-offset variable.
The cues provide neither target elevation nor identity, so participants must still use the prosthetic percept to complete the search. 
We hold the underlying guidance information constant while using modality-appropriate mappings for each sensory channel. We do not attempt to equate physical cue parameters across modalities, since identical temporal or intensity mappings need not be perceptually equivalent across vision, audition, and touch. The comparison therefore evaluates practical implementations of the same guidance information rather than intrinsic differences between sensory modalities.

Because no commercial retinal implants are currently available and implant-user cohorts are small and heterogeneous, direct evaluation with prosthetic vision end users is impractical for a controlled study of this scale.
Immersive \ac{SPV} provides a controlled testbed for interaction design, enabling repeatable manipulation of visual bandwidth while preserving active head-directed search~\cite{hayes_visually_2003,dagnelie_real_2007,kasowski_immersive_2022}.
Our aim is to compare relative guidance effects under this constrained
visual representation, rather than to reproduce blindness or predict
clinical performance.

We evaluated visual, haptic, and auditory guidance ({\bf Figure~\ref{fig:teaser}}) in 19 sighted participants under $10\times10$ and $20\times20$ biologically motivated \ac{SPV}.
Simulations used \texttt{BionicVisionXR}~\cite{kasowski_immersive_2022} with psychophysically validated models of phosphene appearance~\cite{beyeler_model_2019,hou_axonal_2024}.
We found that all three modalities reduced search time and head movement, while auditory and haptic guidance produced approximately 25\% faster overall search than visual guidance.
Trajectory analysis partitioned search into target acquisition, defined as first entry of the target center into the horizontal prosthetic \ac{FOV}, and the subsequent interval to response.
Guidance accelerated acquisition, reduced unnecessary scanning, and also shortened post-acquisition search, with the balance between these benefits depending on cue implementation.

This paper makes three contributions:
\begin{enumerate}[topsep=0pt,itemsep=-1ex,partopsep=0pt,parsep=1ex,
                  leftmargin=14pt,label=\roman*.]
    \item To our knowledge, we present the first within-subject comparison
    of visual, auditory, and haptic out-of-view guidance under immersive
    SPV based on a psychophysically validated phosphene model, with all
    modalities encoding the same underlying target-direction variable.

    \item We partition search into target acquisition and post-acquisition intervals, showing benefits of all cues for acquisition and of the tested haptic and auditory cues for subsequent completion.

    \item We show that the tested auditory and haptic cues outperformed the visual cue under severe visual constraints, while guidance produces substantially tighter final alignment in the coarser $10\times10$ \ac{SPV} condition, including improved localization along the uncued elevation dimension.
\end{enumerate}

\section{Related Work}

\subsection{Object Search and Prosthetic Vision}

Assistive object search places additional demands on an already limited visual channel.
Object search is an important accessibility problem for people who are blind, requiring both spatial localization of a target and sufficient perceptual information to identify it ~\cite{yi_finding_2013,liuObjectFinderOpenVocabularyAssistive2026,singhAssistingBlindReach2026, turkstra_assistive_2025}.
For people with residual vision, systems such as CueSee have used visual highlighting to facilitate target search~\cite{zhao_cuesee_2016}, while broader systems such as SeeingVR modify contrast, magnification, edges, and object appearance to improve access to virtual environments~\cite{zhaoSeeingVRSetTools2019}.

For prosthetic vision, these demands are compounded by the small \ac{FOV} and low spatial resolution of the percept.
For example, Argus~II subtends only about $11^\circ\times19^\circ$~\cite{bloch_argus_2019}, requiring users to scan the scene with head movements.
One line of research therefore focuses on improving the information represented through the prosthesis.
Image simplification, semantic segmentation, and depth- or task-dependent filtering can preserve selected scene information under limited visual bandwidth~\cite{vergnieux_simplification_2017,
sanchez-garcia_semantic_2020,rasla_relative_2022,han_deep_2021}.
These approaches primarily address what should be shown once relevant content falls within the prosthetic \ac{FOV}.

A complementary approach is to guide the user toward relevant content.
Parikh et al.~\cite{parikh_performance_2013} used visual saliency cues during tabletop search under \ac{SPV} and found reductions in search time, head movement, and error.
More recently, Christie et al.~\cite{christie_autonomous_2026} combined prosthetic vision with auditory and haptic feedback for navigation, finding substantial benefits in simulation and similar trends in a case study with a user of the Argus~II epiretinal prosthesis.

Immersive SPV also differs in how faithfully the prosthetic percept is
modeled.
Many prior simulations approximate phosphenes as regular, independent
points of light~\cite{dagnelie_real_2007,vergnieux_simplification_2017}, whereas clinically reported percepts vary in size,
shape, and orientation with retinal location and stimulation
parameters~\cite{beyeler_model_2019,sinclair_appearance_2016}.
\texttt{BionicVisionXR}~\cite{kasowski_immersive_2022} was developed to support head-directed SPV tasks in immersive VR using the psychophysically
validated axon-map model~\cite{beyeler_model_2019,hou_axonal_2024}.
Their work showed that behavioral performance can depend substantially on the assumed phosphene model, motivating its use when
drawing conclusions about interaction under prosthetic vision.
We therefore use \texttt{BionicVisionXR} here to preserve both active
head-directed search and biologically grounded phosphene appearance.

\subsection{Out-of-View Guidance in VR and AR}

Guiding users toward out-of-view content has a long history in \ac{HCI}, \ac{VR}, and \ac{AR}.
Early techniques such as Halo and Wedge encoded the direction and distance of off-screen objects at display boundaries~\cite{baudisch_halo_2003,gustafson_wedge_2008}.
HMD-specific approaches subsequently adapted this problem to immersive environments, including EyeSee360 and other edge-, arrow-, radar-, and focus-plus-context techniques~\cite{gruenefeld_eyesee360_2017,gruenefeld_comparing_2019,krausComparativeStudyOrientation2020,quinn_augmented_2024}.
Comparative studies show that guidance design affects acquisition time, localization accuracy, and the way users move their head while searching~\cite{bork_towards_2018}.
Related work on immersive wayfinding likewise found that directly actionable directional guidance outperformed richer map and compass representations under demanding navigation conditions \cite{varshney_actionable_2026}.

Restricted \ac{FOV} makes these costs more consequential.
Ren et al.~\cite{ren_evaluating_2016} used mixed-reality simulation in \ac{VR} to show how display \ac{FOV} affects task performance and head movement, while Bork et al.~\cite{bork_towards_2018} explicitly analyzed head-rotation trajectories during search with limited-FOV HMDs.
At a larger spatial scale, Kumaran et al.~\cite{kumaran_impact_2023} found that visual navigation aids improved target search in wide-area \ac{AR}, but could also reduce awareness of the surrounding physical environment.
Across these studies, guidance reduces spatial search but can also redirect visual attention away from surrounding content.

Prosthetic vision imposes a much stronger display constraint.
Most immersive out-of-view techniques assume that a high-resolution visual indicator can be added to the user's current view.
Under prosthetic vision, the indicator itself must survive the same severe spatial bottleneck as the scene.
This raises the design question of whether moving guidance outside the visual channel can benefit search under such constraints.

\subsection{Cross-Modal Guidance Under Visual Constraints}

Out-of-view guidance need not be visual.
Auditory and vibrotactile cues have been used to redirect spatial attention without adding content to the visual display~\cite{biocca_attention_2006,stratmann_exploring_2018}.
Recent \ac{VR} work further shows that the benefit of multisensory attentional cues depends on task demands: Jeong et al.~\cite{jeong_differential_2024} found that tactile and visuotactile cues could improve visual-search performance, with multisensory benefits increasing under greater cognitive load.
Cue effectiveness therefore depends on the perceptual demands of the task.

Studies of limited-FOV \ac{AR} make this tradeoff particularly clear.
Marquardt et al.~\cite{marquardt_comparing_2020} compared EyeSee360 with audio-tactile guidance for locating out-of-view targets.
Visual guidance produced faster search, whereas audio-tactile guidance better preserved situation awareness.
Other work has shown that visual guidance can itself incur costs through clutter or occlusion, and that reducing visual assistance can sometimes improve target-search performance~\cite{gruenefeld_improving_2019}.
The preferred modality depends on what the visual channel must simultaneously represent.

Visual prostheses provide an extreme test of this tradeoff.
A visual guidance cue occupies the same sparse phosphene representation used to inspect the target, whereas auditory and haptic cues do not.
In prior cross-modal comparisons, modality is often coupled to differences in cue design or encoding~\cite{marquardt_comparing_2020,christie_autonomous_2026}.
We therefore compare visual, auditory, and haptic cues that encode the same horizontal target-offset variable using modality-specific implementations.
By comparing these cue implementations before and after target acquisition, we test how cross-modal guidance performs as visual bandwidth becomes scarce.

\section{Methods}

\subsection{Participants}

Nineteen participants (11 female, 8 male; ages 19--27 years) were recruited from the Anonymous University community.
All reported normal or corrected-to-normal vision and no known visual or neurological impairments.
Prior \ac{VR} experience was limited: three participants had never used \ac{VR}, 12 had used it once or twice, three reported occasional use ($<$1 time/month), and one reported regular use (1--3 times/month).

All participants completed the experiment and were included in the analysis.
The study was approved by the Anonymous University IRB.
Participants provided written informed consent and received course credit or monetary compensation.

\subsection{Apparatus}

The experiment was implemented in Unity and presented on an HTC VIVE Pro Eye \ac{HMD} with six-degree-of-freedom head tracking.
Participants sat in a swivel chair and could rotate their head and upper body freely.
Tracked VIVE controllers were used for responses and delivered vibrotactile feedback in the haptic condition.

The application ran on a desktop equipped with an NVIDIA GeForce RTX~3080 GPU.
Head pose and controller state were recorded continuously.

\begin{figure}[!t]
    \centering
    \includegraphics[width=\linewidth]{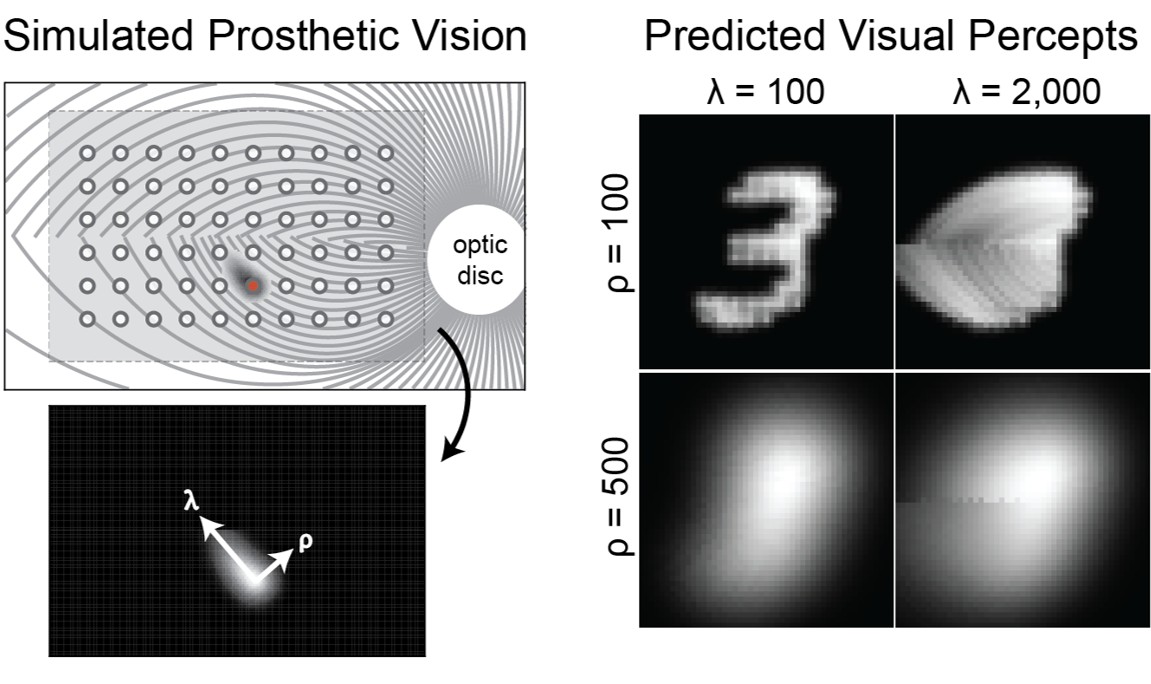}
    \caption{Axon map model for simulated prosthetic vision. \emph{Left:} Electrical stimulation (red disc) of a nerve fiber bundle in the retina (gray lines) leads to elongated tissue activation (gray shaded region) and phosphenes whose shape can be described by two parameters, $\lambda$ (axonal spread) and $\rho$ (radial spread). \emph{Right:} Example percepts for an MNIST digit for varing $\rho$ and $\lambda$ values. 
    In the experiment, we used $\rho=\SI{400}{\micro\meter}$ for the $10\times10$ condition and $\rho=\SI{200}{\micro\meter}$ for the $20\times20$ condition, with $\lambda=\SI{100}{\micro\meter}$ in both conditions.}
    \label{fig:spv}
\end{figure}

\subsection{Simulated Prosthetic Vision}
\label{sec:spv}

Retinal implants electrically stimulate retinal neurons, producing spots or streaks of light called \emph{phosphenes}.
Activating multiple electrodes therefore produces a spatial phosphene pattern, which we refer to as the prosthetic \emph{percept} ({\bf Figure~\ref{fig:spv}}).

We used \texttt{BionicVisionXR}~\cite{kasowski_immersive_2022} to simulate epiretinal stimulation with the axon-map model~\cite{beyeler_model_2019,hou_axonal_2024}.
At each frame, the VR camera image was sampled at simulated electrode locations to determine electrode activation.
The model predicted the resulting percept, with $\rho$ controlling radial phosphene spread and $\lambda$ controlling elongation along retinal nerve-fiber pathways ({\bf Figure~\ref{fig:spv}}).
Predicted brightness at visual-field location $(r,\theta)$ was
\begin{equation}
    b(r,\theta)
    =
    \max_{p \in R(\theta)}
    \sum_{e \in E}
    a_e
    \exp\left(
    -\frac{d_e^2}{2\rho^2}
    -\frac{d_{\mathrm{soma}}^2}{2\lambda^2}
    \right),
    \label{eq:axon-map}
\end{equation}
where $a_e$ is electrode activation, $d_e$ is distance to electrode $e$, and $d_{\mathrm{soma}}$ is distance along the corresponding nerve-fiber pathway.

\begin{figure}[!t]
    \centering
    \includegraphics[width=\linewidth]{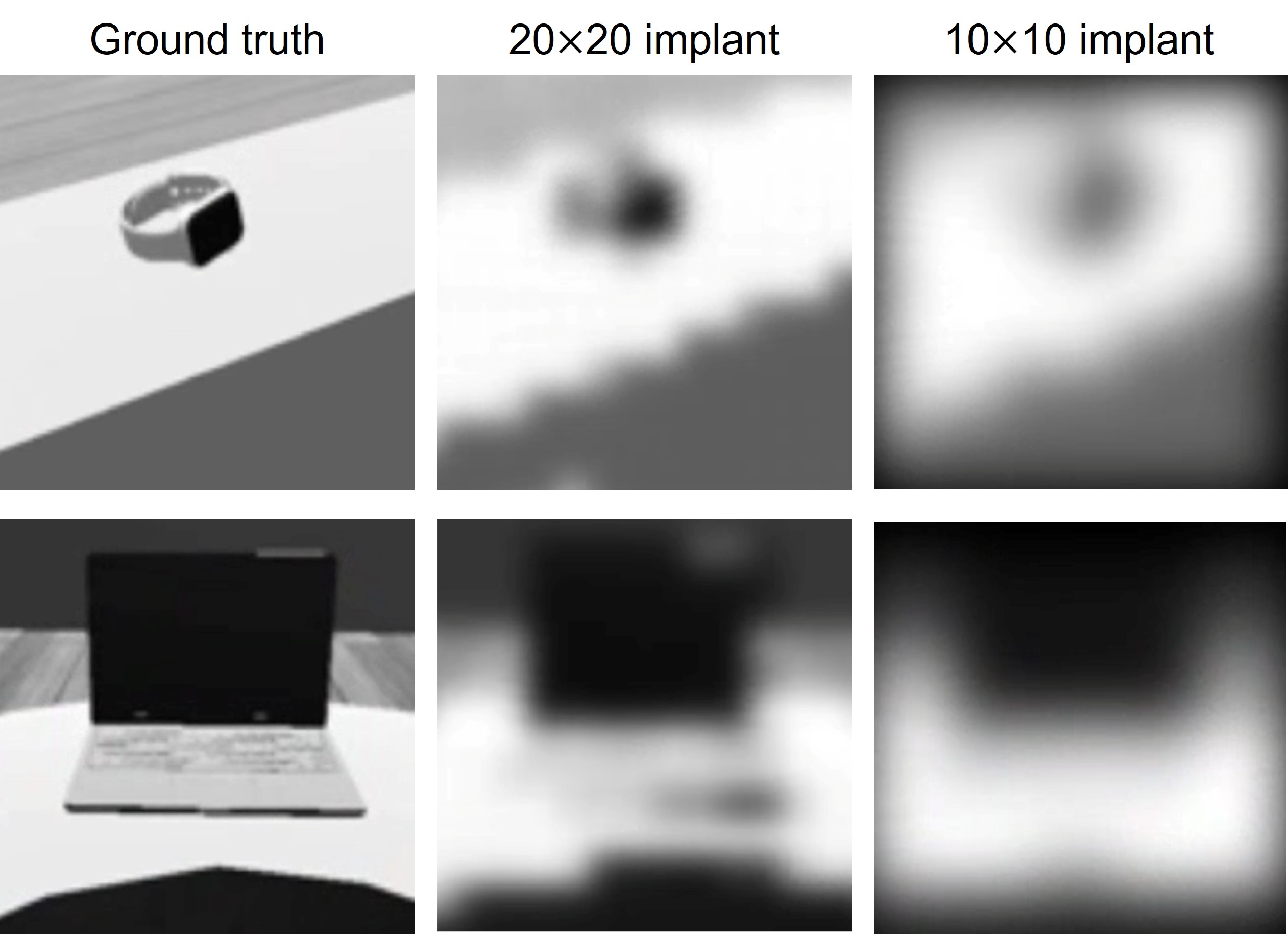}
    \caption{
        Representative scene views under natural vision (left) and simulated
        prosthetic vision with $20\times20$ (center) and $10\times10$ (right)
        electrode arrays.
        Both prosthetic conditions span the same $30^\circ\times30^\circ$
        visual field, while electrode density and phosphene spread differ as
        described in Sec.~\ref{sec:spv}.
        Views were captured separately from approximately matched viewpoints and
        are intended to illustrate perceptual appearance rather than pixel-wise
        correspondence.
    }
    \label{fig:methods-scene}
\end{figure}

We compared $10\times10$ and $20\times20$ electrode arrays spanning the same $4.5\times\SI{4.5}{\milli\meter}$ retinal area and $30^\circ\times30^\circ$ visual field.
Electrode spacing was \SI{500}{\micro\meter} and approximately \SI{237}{\micro\meter}, respectively, with locations mapped to visual angle using the Watson retinal map~\cite{watson_formula_2014}.
To maintain similar phosphene overlap across electrode spacings, we used $\rho=\SI{400}{\micro\meter}$ for the $10\times10$ condition and $\rho=\SI{200}{\micro\meter}$ for the $20\times20$ condition.
We fixed $\lambda=\SI{100}{\micro\meter}$ in both conditions to limit axon-aligned elongation and hold phosphene shape constant across conditions.
Thus, the two SPV conditions differed jointly in electrode density and radial phosphene spread while retinal extent, visual-field extent, and $\lambda$ were held fixed.
Representative percepts are shown in {\bf Figure~\ref{fig:methods-scene}}.

These conditions were intended to represent substantially different prosthetic-resolution regimes rather than small changes in display resolution.
A fourfold change in electrode count is substantial in visual prosthesis design, comparable in scale to generational changes across systems such as Argus~I~\cite{humayun_pattern_1999}, Argus~II~\cite{bloch_argus_2019}, and PRIMA~\cite{holz_subretinal_2025}.

Image intensity was normalized to $[0,1]$, sampled at the electrode locations, and scaled by 0.25 before percept rendering.
For real-time rendering in Unity, we exported the PyTorch axon-map implementation to \ac{ONNX} and executed it on the GPU using Unity Sentis.
The predicted percept was composited with the camera view on each frame.
We verified the exported pipeline against the native PyTorch implementation using identical inputs and both experimental parameter settings, confirming equivalent predicted percepts.

\subsection{Task and Environment}

Participants searched for specified objects in a virtual workspace while viewing the scene exclusively through \ac{SPV}.
The environment contained a semicircular desk surrounding the seated participant ({\bf Figure~\ref{fig:teaser}B}).

\begin{figure}[!t]
    \centering
    \includegraphics[width=\linewidth]{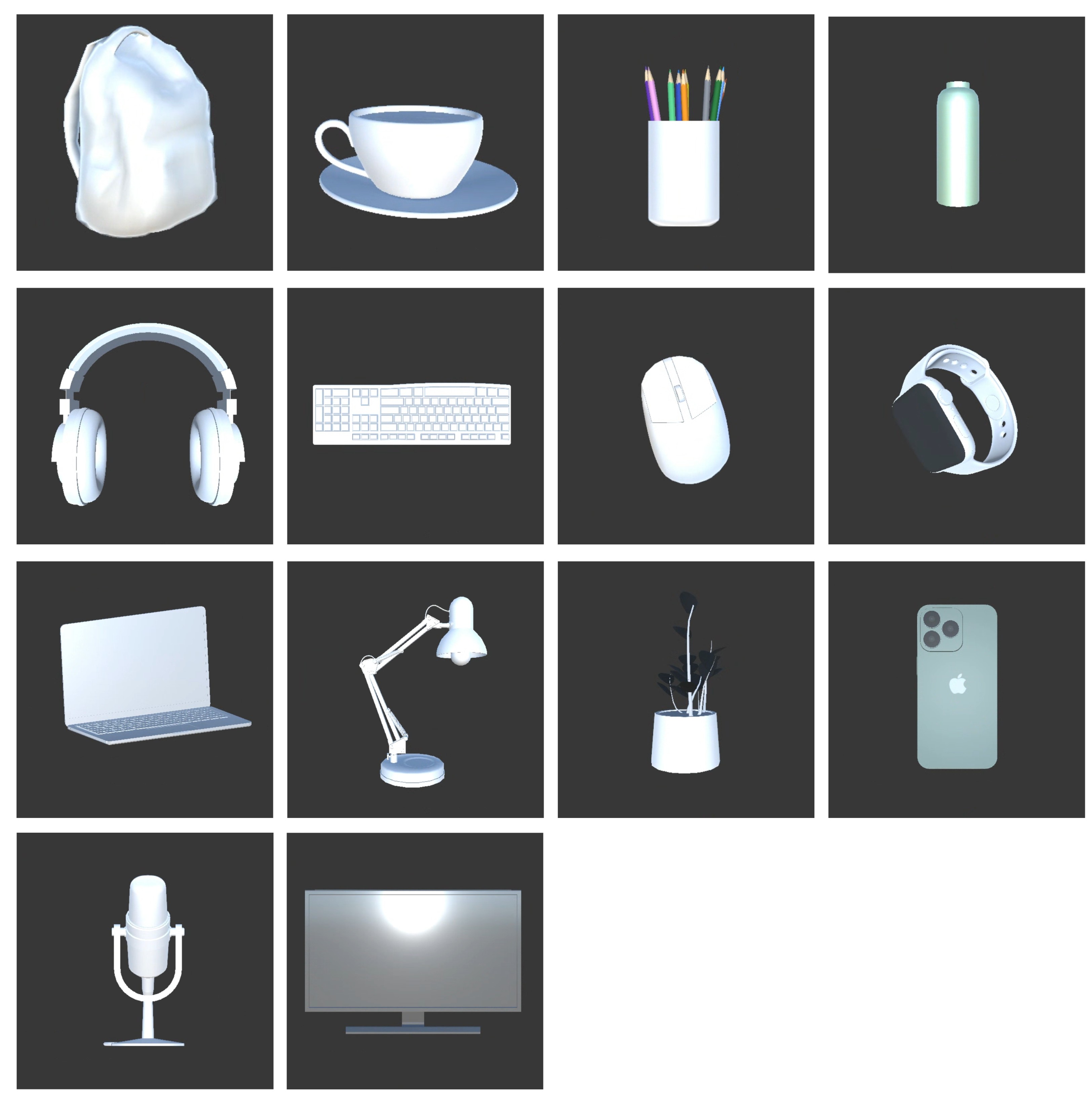}
    \caption{The 14 target objects used in the search task.}
    \label{fig:methods-objects}
\end{figure}

The target set contained 14 familiar desk and household objects spanning a range of sizes and coarse shapes ({\bf Figure~\ref{fig:methods-objects}}).
Objects could appear at eight predefined desk locations spanning
approximately $200^\circ$ of horizontal azimuth ($-100^\circ$ to
$+100^\circ$).
In the single-object condition, only the target was present; in the cluttered condition, the target appeared with four distractors.

Guidance and SPV condition were blocked, yielding eight 20-trial blocks.
Each block contained 10 single-object and 10 cluttered trials drawn from a pre-generated trial set specifying target identity, target location, and distractor configuration.
We included clutter as a secondary manipulation of scene complexity to test whether guidance effects generalized from isolated targets to search among distractors.

Camera height was varied across trials by -0.5, 0, +0.5, or +\SI{1.0}{\meter} relative to its nominal position so that horizontal guidance alone was insufficient and participants had to localize the target vertically using \ac{SPV}.
The same offset was prohibited on consecutive trials.

The prosthetic \ac{FOV} spanned $30^\circ\times30^\circ$.
Because target positions extended around the semicircular workspace, targets typically began outside the horizontal prosthetic \ac{FOV}.
Participants therefore had to rotate their head to bring the target into the represented region before completing the search.

We define \emph{azimuthal acquisition} as the first frame in which the target center entered the $\pm15^\circ$ horizontal prosthetic \ac{FOV}.
We used target center so acquisition time would not depend on object size.
Note that acquisition marks first entry only: the target could subsequently leave and re-enter the \ac{FOV} while participants localized it
vertically and distinguished it from distractors using \ac{SPV}.


\subsection{Guidance Conditions}

All guidance conditions encoded the horizontal angular offset $\theta$ between the participant's head direction and the target azimuth.
Cue laterality indicated whether the target lay to the left or right, and cue urgency increased as $|\theta|$ decreased.
Within $\pm5^\circ$ of horizontal alignment, cues switched to a centered
state. This threshold applied only to target azimuth and provided no
information about vertical target position. Therefore, participants still had to rely on \ac{SPV} to complete the search.
The cues encoded the same underlying directional variable using modality-specific mappings described below.
These mappings were designed to make target direction and increasing alignment readily actionable within each modality, rather than to impose identical physical transfer functions across modalities. Cross-modal equivalence in salience or temporal discriminability was not assumed.

\paragraph{Baseline.}
No directional guidance was provided.

\paragraph{Visual.}
Pulsating cues appeared at the left and right edges of the prosthetic display ({\bf Figure~\ref{fig:teaser}C}).
For target offsets greater than $5^\circ$, only the cue on the corresponding side was active; within $\pm5^\circ$, both cues pulsated.
Cue visibility toggled at intervals ranging from 0.05 to \SI{1.0}{\second}, increasing linearly with $|\theta|$ over a $270^\circ$ range.
The cues were rendered through the same \ac{SPV} pipeline as the scene.
Guidance remained active after azimuthal acquisition, so the visual cue continued to be rendered within the prosthetic percept while participants completed the search.

\paragraph{Audio.}
Stereo-panned beeps indicated target direction: targets more than $5^\circ$ from alignment produced a fully left- or right-panned beep, whereas offsets within $\pm5^\circ$ produced a centered beep.
The inter-beep interval increased from 0.05~s at alignment to \SI{1.0}{\second} at offsets of $135^\circ$ or greater.

\paragraph{Haptic.}
Vibrotactile pulses were delivered through the controller corresponding to target direction; both controllers vibrated within $\pm5^\circ$ of alignment.
The inter-pulse interval increased from 0.08 to \SI{0.60}{\second} over $0$--$135^\circ$ of target offset.
Pulse intensity decreased from 0.30 to 0.10 and duration from 0.15 to \SI{0.05}{\second} over the same range; vibration frequency was \SI{150}{\hertz}.

\subsection{Experimental Design and Procedure}

We used a within-subjects $4\times2\times2$ design with guidance (baseline, visual, haptic, audio), SPV condition ($20\times20$, $10\times10$), and scene clutter (one vs. five objects).
Clutter served as a secondary factor testing the robustness of guidance effects to distractor presence.

Each participant completed 160 experimental trials in eight 20-trial guidance $\times$ SPV-condition blocks.
The trials were specified in advance as eight 20-trial lists, each containing 10 single-object and 10 cluttered trials with predefined target identity, target location, and distractor configuration.
Across the full 160-trial set, each of the eight target locations occurred 20 times, equally divided between single-object and cluttered trials.

Guidance order followed a four-condition balanced Latin square, while SPV-condition order ($20\times20$ first vs.\ $10\times10$ first) was counterbalanced independently, yielding eight block-order sequences.
Participants completed all four guidance conditions at one SPV condition before switching to the other, with the same guidance order repeated.
Ten participants completed the $20\times20$ condition first and nine completed the $10\times10$ condition first.

Participants first completed 10 practice trials in a separate but structurally similar environment under normal vision so they could familiarize themselves with the guidance cues.
At the start of each experimental trial, the target name was displayed
as text (e.g., ``Find: Mouse'').
The search scene then appeared under the assigned \ac{SPV} and guidance condition.
Participants searched by rotating their head and upper body and were instructed to center the target in their prosthetic \ac{FOV} before indicating that they had located it using the trigger button on either VIVE controller.
Participants could respond before target-center entry or after the target center had subsequently left the horizontal \ac{FOV}.

After each block, participants rated search difficulty on a 1--10 scale.
At the end of the experiment, participants identified the guidance methods they found most and least helpful.

\subsection{Measures}

\paragraph{Search time.}
Search time was measured from scene onset to the participant's response.

\paragraph{Azimuthal acquisition latency.}
For trials beginning with the target center outside the horizontal prosthetic \ac{FOV}, acquisition latency was measured from scene onset to first target-center entry into the $\pm15^\circ$ horizontal \ac{FOV}.
This marks entry into the represented region, not the instructed response criterion of centering the target.

\paragraph{Post-acquisition elapsed time.}
For initially out-of-view trials with observed acquisition, post-acquisition elapsed time ($T_{\mathrm{post}}$) was the interval from first target-center entry into the horizontal prosthetic \ac{FOV} ($T_{\mathrm{acq}}$) to the participant's response ($T_{\mathrm{response}}$): $T_{\mathrm{post}} = T_{\mathrm{response}} - T_{\mathrm{acq}}$.
Thus, search time from scene onset to response was partitioned as $T_{\mathrm{search}} = T_{\mathrm{acq}} + T_{\mathrm{post}}$.

\paragraph{Final head-target angular offset.}
Final head-target angular offset was the 3D angular separation at response between the head-forward direction and the direction from the tracked head position to the target center, computed from the normalized vector dot product.
It therefore measures how closely the participant's head was oriented toward the target when they reported finding it.
Because the prosthetic \ac{FOV} was square in azimuth-elevation coordinates, a target could lie within $\pm15^\circ$ in both dimensions while its three-dimensional angular separation from head-forward exceeded $15^\circ$.

\paragraph{Total head rotation.}
Total head rotation was the accumulated three-dimensional angular change in head orientation over the complete trial, rather than yaw alone.

\paragraph{Excess yaw to acquisition.}
For initially out-of-view trials with observed acquisition, the minimum yaw required to bring the target center to the horizontal \ac{SPV} boundary was $d_{\min}=\max\left(|\theta_0|-15^\circ,\,0\right)$,
where $\theta_0$ was the target-center azimuth at trial onset.
Excess yaw was the accumulated absolute head-yaw rotation before acquisition minus this minimum required rotation (clipped at zero).
Values near zero indicate a geometrically direct orienting trajectory.

\paragraph{Response alignment.}
Because the trigger response did not require participants to select an object, object-identification accuracy was not recorded directly. We therefore also recorded whether the target center lay within the $30^\circ\times30^\circ$ SPV field of view at response $\pm15^\circ$ in azimuth and elevation).

\paragraph{Subjective measures.}
After each block, participants rated search difficulty on a 10-point scale (1 = very easy, 10 = very difficult).
After the experiment, they ranked the three guidance modalities by perceived helpfulness and answered open-ended questions about search strategy, cue usability, visual resolution, clutter, and task difficulty.
Open-ended responses were summarized descriptively rather than subjected to formal thematic analysis.

\begin{figure*}[!t]
    \centering
    \includegraphics[width=\linewidth]{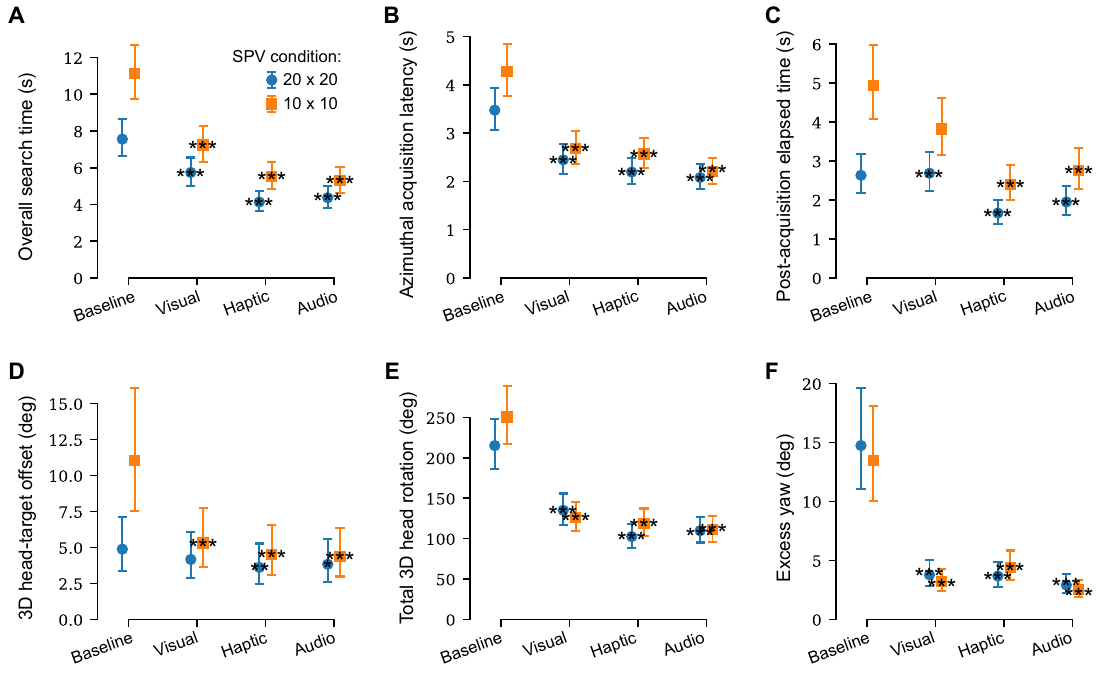}
    \caption{
    Effects of guidance modality and prosthetic resolution on object search.
    Points show back-transformed marginal estimates from the mixed-effects
    models; error bars denote 95\% confidence intervals.
    \textbf{A:} Overall search time.
    \textbf{B:} Azimuthal acquisition latency, measured from trial onset
    until the target center first entered the horizontal prosthetic field
    of view.
    \textbf{C:} Post-acquisition elapsed time, measured from first target-center entry until response.
    \textbf{D:} Final head-target offset at response.
    \textbf{E:} Total accumulated head rotation over the trial.
    \textbf{F:} Excess yaw before acquisition, defined as accumulated yaw
    beyond the minimum rotation geometrically required to bring the target
    center into the prosthetic \ac{FOV}.
    Panels B, C, and F include trials with observed target-center entry.
    Estimates are averaged over scene clutter at the mean continuous
    covariates.
    Asterisks indicate only the planned Holm-corrected
    guidance-versus-baseline contrasts within each SPV condition
    ($*p<.05$, $**p<.01$, $***p<.001$); other planned contrasts are reported in the main text.
    }
    \label{fig:results}
\end{figure*}

\subsection{Statistical Analysis}

Trial-level outcomes were analyzed using linear mixed-effects models in Python with \texttt{statsmodels}.
Guidance, SPV condition, and clutter were categorical fixed effects with all interactions, using sum-to-zero contrasts. Standardized initial target eccentricity and trial order were included as covariates. All 19 participants completed all 160 trials ($N=3{,}040$), and no trials were excluded as outliers.

Search time and final head-target offset were log-transformed; total head rotation was analyzed as $\log(1+x)$. Models included random intercepts for participant and participant-by-block, with an additional crossed random intercept for target object where supported by the data. Models were fit by restricted maximum likelihood.

Fixed effects were evaluated using omnibus Wald $\chi^2$ tests. The seven categorical terms across the six primary outcome models in Tables~\ref{tab:omnibus} and~\ref{tab:trajectory_omnibus} constituted a family of 42 factorial tests; Holm-adjusted $p$ values for this family are reported in Table~S1 in the Supplemental Materials. Planned contrasts compared each guidance condition with baseline within each SPV condition, pooled auditory and haptic guidance with visual guidance, and tested whether guidance effects differed across SPV conditions. Contrasts were Holm-corrected within their prespecified families. Log-scale estimates were exponentiated for interpretation as ratios or percentage changes.

\paragraph{Trajectory-level analyses.}
Acquisition latency, post-acquisition elapsed time, and excess yaw were analyzed for initially out-of-view trials with observed target-center entry ($n=2{,}870$). Acquisition and post-acquisition times were log-transformed, and excess yaw was analyzed as $\log(1+x)$. These models used the same factorial structure and random effects as the primary models, replacing initial eccentricity with standardized initial out-of-view distance $d_{\min}$.

Accumulated in-view dwell time, post-acquisition time outside the horizontal \ac{FOV}, and re-exits after first acquisition were treated as secondary target-maintenance measures; detailed results are reported in Table~S3 in the Supplemental Materials.

Because 112 of the 2,982 initially out-of-view trials ended before target-center entry, acquisition latency was not observed on those trials. We therefore performed a sensitivity analysis using a log-normal accelerated failure-time model~\cite{wei_accelerated_1992}, which incorporates both observed acquisition times and trials that ended before acquisition. The model used the same fixed effects and covariates as the primary acquisition model and participant-clustered standard errors. Effects are reported as acquisition-time ratios, with values below 1 indicating faster acquisition; full results are reported in Table~S2 in the Supplemental Materials.

\paragraph{Response alignment.}
Because acquisition before response was itself condition-dependent, we
also analyzed whether target-center acquisition occurred before response
on the 2,982 initially out-of-view trials using a participant-clustered
binomial GEE.
Post-acquisition elapsed time and excess yaw are conditional process
measures defined only for trials with observed acquisition and therefore
should not be interpreted as marginal effects over all trials.
As a sensitivity analysis, we repeated these models on the stricter
subset in which target-center acquisition was observed and the target
center was within the horizontal SPV field of view at response.
Response alignment at the time of the trigger response was analyzed
separately using a participant-clustered binomial GEE.

\paragraph{Difficulty ratings.}
Block-level difficulty ratings were analyzed with a mixed-effects model including guidance, SPV condition, and their interaction, with a random intercept for participant. Planned within-SPV-condition guidance contrasts were Holm-corrected.

\section{Results}

\subsection{Effects of Guidance on Search Performance}

All three guidance modalities substantially reduced search time ({\bf Figure~\ref{fig:results}A}; {\bf Table~\ref{tab:omnibus}}).
At $20\times20$, search time decreased by 24\% with visual guidance, 45\% with haptic guidance, and 42\% with audio guidance; at $10\times10$, the corresponding reductions were 35\%, 50\%, and 52\% (all Holm-corrected $p<.001$).
Pooling auditory and haptic conditions, nonvisual guidance produced approximately 25\% faster search than visual guidance in both SPV conditions.

Search time was longer in the $10\times10$ than the $20\times20$ SPV condition ($\chi^2(1)=75.71$, $p<.001$). There was no reliable guidance $\times$ SPV-condition interaction ($\chi^2(3)=5.33$, $p=.149$): the advantage of nonvisual over visual guidance was similar across the two SPV conditions.

\begin{table*}[!th]
\centering
\caption{
Omnibus Wald tests for whole-trial outcomes under sum-to-zero contrast
coding. Lower-order categorical effects therefore represent effects
averaged over levels of interacting factors.
Search time and final head-target offset were log-transformed; total
head rotation was analyzed as $\log(1+x)$.
Initial target eccentricity and trial order were standardized covariates.
Values are $\chi^2(\mathrm{df})$; raw omnibus $p$ values are shown here, with Holm-adjusted values across the 42 factorial tests from Tables~1--2 reported in Table~S1 in the Supplemental Materials.
}
\label{tab:omnibus}
\begin{tabular}{lrr rr rr}
\hline
& \multicolumn{2}{c}{Search time}
& \multicolumn{2}{c}{Head-target offset}
& \multicolumn{2}{c}{Head rotation} \\
Effect
& $\chi^2(\mathrm{df})$ & $p$
& $\chi^2(\mathrm{df})$ & $p$
& $\chi^2(\mathrm{df})$ & $p$ \\
\hline
SPV condition
    & 75.71 (1)  & $<.001$
    & 57.73 (1)  & $<.001$
    & 2.56 (1)   & $.110$ \\

Guidance
    & 284.94 (3) & $<.001$
    & 109.99 (3) & $<.001$
    & 262.08 (3) & $<.001$ \\

Clutter
    & 0.00 (1)   & $.986$
    & 41.31 (1)  & $<.001$
    & 3.39 (1)   & $.066$ \\

SPV condition $\times$ Guidance
    & 5.33 (3)   & $.149$
    & 34.23 (3)  & $<.001$
    & 6.25 (3)   & $.100$ \\

SPV condition $\times$ Clutter
    & 0.62 (1)   & $.432$
    & 4.58 (1)   & $.032$
    & 0.02 (1)   & $.877$ \\

Guidance $\times$ Clutter
    & 5.31 (3)   & $.150$
    & 52.14 (3)  & $<.001$
    & 12.07 (3)  & $.007$ \\

SPV condition $\times$ Guidance $\times$ Clutter
    & 7.38 (3)   & $.061$
    & 9.25 (3)   & $.026$
    & 3.73 (3)   & $.292$ \\

\hline
\end{tabular}
\end{table*}

\subsection{Target Acquisition and Scanning Behavior}

Guidance strongly accelerated target acquisition
({\bf Figure~\ref{fig:results}B}; {\bf Table~\ref{tab:trajectory_omnibus}}).
All three modalities were faster than baseline in both SPV conditions
(all Holm-corrected $p<.001$).
Pooling auditory and haptic conditions, nonvisual guidance was 12.7\%
faster than visual guidance at $20\times20$ ($p=.017$) and 11.3\%
faster at $10\times10$ ($p=.020$).
Acquisition was also slower in the $10\times10$ condition
($\chi^2(1)=18.45$, $p<.001$), with no guidance $\times$ SPV-condition
interaction ($\chi^2(3)=3.72$, $p=.293$).

Of 3,040 trials, 2,982 (98.1\%) began with the target center outside the
horizontal prosthetic FOV.
Acquisition before response was condition-dependent
(Table~S2): it occurred on 658/730 baseline trials (90.1\%), compared
with 732/750 visual (97.6\%), 737/752 haptic (98.0\%), and 743/750
audio trials (99.1\%).
The difference was largest at $10\times10$, where acquisition occurred
before response on 85.3\% of baseline trials versus 96.8--98.9\% of
guided trials.
The censoring-aware AFT analysis therefore included all 2,982
initially out-of-view trials when estimating acquisition latency
(Table~S2), whereas subsequent trajectory measures are conditional on
observed acquisition.

Guidance also increased the likelihood that the target center was within the full SPV FOV at response.
This occurred on 87.9\% of baseline trials versus 93.7--96.6\% of guided
trials at $20\times20$, and on 65.3\% versus 86.1--91.6\% at
$10\times10$ (all planned guidance-versus-baseline contrasts,
Holm-corrected $p<.001$).

\begin{figure*}[!th]
    \centering
    \includegraphics[width=\linewidth]{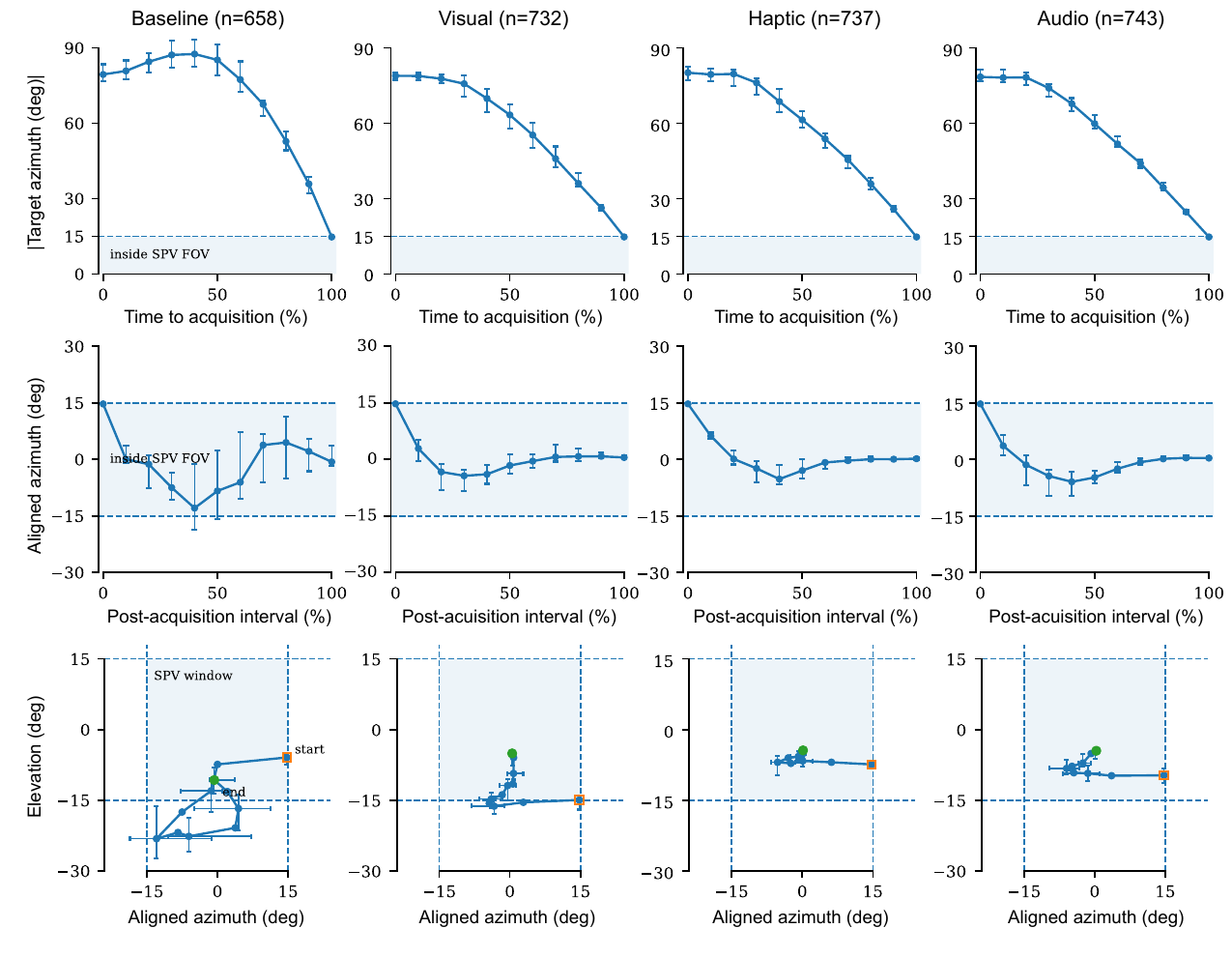}
    \caption{
        Target-relative head-orientation trajectories by guidance condition.
        \emph{Top:} horizontal target eccentricity from trial onset to azimuthal acquisition,
        defined as first target-center entry into the $\pm15^\circ$ horizontal SPV
        \ac{FOV}.
        \emph{Middle:} acquisition-side-aligned azimuth from acquisition to response.
        \emph{Bottom:} the same post-acquisition interval in aligned azimuth--elevation space.
        Trajectories include initially out-of-view trials with observed target-center
        entry; the resulting trial counts are shown for each guidance condition.
        Time was normalized within each interval; trajectories were averaged within
        participant and condition, pooling SPV conditions and clutter, then summarized
        across participants by the median and interquartile range.
        Shading marks the $30^\circ\times30^\circ$ SPV \ac{FOV}.
    }
    \label{fig:head-trajectories}
\end{figure*}

Guidance also made orienting substantially more direct
({\bf Figure~\ref{fig:results}F}; {\bf Figure~\ref{fig:head-trajectories}}).
Excess yaw fell by 67--81\% across modalities and SPV conditions
(all Holm-corrected $p<.001$), with no guidance $\times$ SPV-condition
interaction ($\chi^2(3)=2.66$, $p=.447$).
This result was unchanged in the stricter sensitivity subset used for
the post-acquisition analysis.

Across the complete trial, guidance similarly reduced total head
rotation ({\bf Figure~\ref{fig:results}E}).
Visual, haptic, and audio guidance reduced rotation by 37\%, 52\%, and
49\% at $20\times20$, and by 49\%, 52\%, and 56\% at $10\times10$
(all Holm-corrected $p<.001$).

\begin{table*}[!th]
\centering
\caption{
Omnibus Wald tests for trajectory-level mixed-effects models under
sum-to-zero contrast coding.
Lower-order categorical effects therefore represent effects averaged
over levels of interacting factors.
Acquisition latency and post-acquisition elapsed time were
log-transformed; excess yaw was analyzed as $\log(1+x)$.
Models included initially out-of-view trials with observed target-center
entry ($n=2{,}870$), with standardized initial out-of-view distance and
trial order as covariates.
Values are $\chi^2(\mathrm{df})$; raw omnibus $p$ values are shown here,
with Holm-adjusted values across the 42 factorial tests from Tables~1--2
reported in Table~S1 in the Supplemental Materials.
}
\label{tab:trajectory_omnibus}
\begin{tabular}{lrr rr rr}
\hline
& \multicolumn{2}{c}{Acquisition latency}
& \multicolumn{2}{c}{Post-acquisition time}
& \multicolumn{2}{c}{Excess yaw} \\
Effect
& $\chi^2(\mathrm{df})$ & $p$
& $\chi^2(\mathrm{df})$ & $p$
& $\chi^2(\mathrm{df})$ & $p$ \\
\hline
SPV condition
& 18.45 (1) & $<.001$
& 150.71 (1) & $<.001$
& 0.34 (1) & $.562$ \\

Guidance
& 216.05 (3) & $<.001$
& 188.27 (3) & $<.001$
& 220.47 (3) & $<.001$ \\

Clutter
& 0.04 (1) & $.845$
& 0.18 (1) & $.671$
& 7.99 (1) & $.005$ \\

SPV condition $\times$ Guidance
& 3.72 (3) & $.293$
& 11.32 (3) & $.010$
& 2.66 (3) & $.447$ \\

SPV condition $\times$ Clutter
& 0.01 (1) & $.935$
& 0.84 (1) & $.360$
& 3.02 (1) & $.082$ \\

Guidance $\times$ Clutter
& 0.19 (3) & $.980$
& 2.02 (3) & $.569$
& 0.96 (3) & $.811$ \\

SPV condition $\times$ Guidance $\times$ Clutter
& 2.67 (3) & $.445$
& 0.30 (3) & $.960$
& 5.43 (3) & $.143$ \\
\hline
\end{tabular}
\end{table*}

\subsection{Post-Acquisition Search}

Post-acquisition elapsed time differed strongly across guidance conditions
({\bf Figure~\ref{fig:results}C}; {\bf Table~\ref{tab:trajectory_omnibus}}).
At $20\times20$, haptic and audio guidance reduced post-acquisition time
by 37\% and 26\%, respectively, whereas visual guidance had little effect.
At $10\times10$, the corresponding reductions were 51\%, 44\%, and 23\%.

Because this measure is defined only for trials with observed acquisition,
we tested robustness in the stricter subset in which the target center was
also within the horizontal \ac{FOV} at response ($n=2{,}763$).
The haptic and auditory benefits remained significant at both SPV
conditions, whereas visual guidance no longer differed reliably from
baseline.
Thus, the post-acquisition benefit was robust for the tested haptic and
auditory cues but not for visual guidance.

Secondary maintenance measures showed that guidance also reduced loss of
target alignment after acquisition (Table~S3 in the Supplemental
Materials).
This helps explain why accumulated in-view dwell is not a direct measure
of post-acquisition processing time: longer dwell can reflect better
maintenance of the target within the \ac{FOV}.

\subsection{Effects of Guidance on Final Head-Target Alignment}

Guidance affected final head-target alignment differently across the two SPV conditions ({\bf Figure~\ref{fig:results}D}; {\bf Table~\ref{tab:omnibus}}). 
Head-target angular offset was larger in the $10\times10$ condition ($\chi^2(1)=57.73$, $p<.001$), with a strong guidance $\times$ SPV-condition interaction ($\chi^2(3)=34.23$, $p<.001$).

At $20\times20$, visual, haptic, and audio guidance reduced final head-target offset by 15\%, 26\%, and 22\%, respectively; 
the haptic ($p=.003$) and audio ($p=.017$) contrasts were significant after correction,
whereas the visual contrast was not ($p=.085$).
At $10\times10$, the reductions increased to 52\%, 59\%, and 61\% (all $p<.001$).
For every modality, this effect was significantly larger at $10\times10$ than at $20\times20$
(all Holm-corrected $p<.001$).

Because the guidance cues directly encoded azimuth but provided no elevation information, we decomposed final alignment into absolute azimuth and elevation errors. Guidance strongly reduced both azimuth and elevation error, with guidance effects on both dimensions depending on SPV condition (Table~S4 in the Supplemental Materials). 
At $20\times20$, guidance did not reliably improve elevation error. At $10\times10$, however, visual, haptic, and audio guidance reduced elevation error by approximately 45\%, 55\%, and 58\%, respectively, with all three contrasts significant after correction (Table~S5).
Thus, although guidance supplied only horizontal target information, under coarse SPV it also improved localization in the uncued vertical dimension.

Final head-target offset also showed a guidance $\times$ clutter interaction (Table~\ref{tab:omnibus}), but the guidance $\times$ SPV condition $\times$ clutter interaction did not survive correction across the omnibus factorial tests. We therefore do not interpret the SPV-dependent guidance effect as being specifically amplified by clutter.

\subsection{Subjective Experience and Search Strategies}

\begin{figure}[!t]
    \centering
    \includegraphics[width=\linewidth]{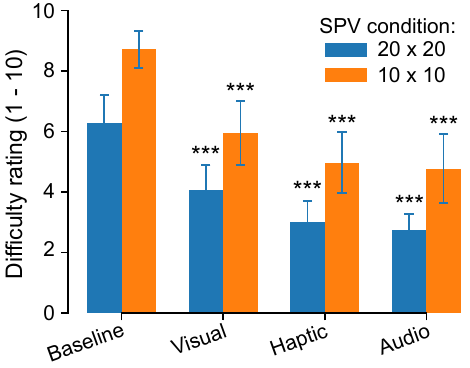}
    \caption{
        Reported search difficulty by guidance modality and SPV condition.
        Bars show participant means with 95\% confidence intervals; higher scores indicate greater difficulty.
        Asterisks mark planned within-SPV-condition contrasts against baseline, Holm-corrected within each contrast family ($***p<.001$).
    }
    \label{fig:difficulty}
\end{figure}

Participants also rated guided search as easier ({\bf Figure~\ref{fig:difficulty}}).
Guidance reduced reported difficulty ($\chi^2(3)=160.35$, $p<.001$) and was higher in the $10\times10$ SPV condition ($\chi^2(1)=79.61$, $p<.001$); the interaction was not significant ($p=.840$).
All guided conditions were rated easier than baseline in both SPV conditions (all Holm-corrected $p<.001$), and 18 of 19 participants
reported that object identification was easier at $20\times20$ than at $10\times10$.

Participants strongly preferred nonvisual guidance.
Haptic was rated most helpful by 9/19 participants and audio by 8/19,
whereas 15/19 rated visual guidance least helpful.

Eleven participants described using guidance to establish horizontal
direction before searching vertically or identifying the object
(``Go til the cue for horizontal, then look up and down.'').
Without guidance, participants described broader scanning, including an
``up and down wave motion ... to sweep the entire plane.''

Nine participants described the visual cue as distracting or confusing,
often because it blended with or obscured scene content.
One participant noted that the dots would ``brighten up and make it hard
to see the objects.''

\section{Discussion}

We studied how visual, auditory, and haptic guidance affect out-of-view object search when the visual display is severely bandwidth-limited and each modality conveys the same directional information.

Guidance accelerated target acquisition, reduced unnecessary head rotation, and continued to improve performance after acquisition, although the balance between these benefits depended on cue implementation ({\bf Figure~\ref{fig:results}B,C,F}). Pooling auditory and haptic conditions, nonvisual guidance produced faster acquisition than visual guidance, while excess yaw did not reliably differ between them. Guidance also produced substantially tighter final head-target alignment in the $10\times10$ than the $20\times20$ SPV condition, including improved localization along the uncued elevation dimension. The tested cue implementations therefore differed not only in overall performance, but in how they supported successive stages of search.

\subsection{Guidance Accelerates Acquisition and Stabilizes Subsequent Search}

Out-of-view guidance is already known to shorten search and alter
head-search behavior in immersive displays
~\cite{bork_towards_2018,gruenefeld_comparing_2019,
krausComparativeStudyOrientation2020,kumaran_impact_2023}.
Parikh et al.~\cite{parikh_performance_2013} reported similar benefits
from visual saliency cues under \ac{SPV}.
We replicate this whole-trial benefit under biologically grounded \ac{SPV}.

Our trajectory analysis shows where this benefit arose.
Without guidance, participants often rotated substantially farther than
was geometrically necessary to acquire the target.
All three guidance modalities removed most of this excess yaw and
produced more direct convergence
({\bf Figure~\ref{fig:results}F}; {\bf Figure~\ref{fig:head-trajectories}}).
The reduction in total head rotation
({\bf Figure~\ref{fig:results}E}) therefore reflects less exploratory scanning,
not simply shorter trial duration.
Unlike prior head-movement analyses under limited-\ac{FOV} guidance~\cite{bork_towards_2018}, excess yaw here was anchored to an explicit acquisition event.

Guidance remained useful after acquisition, but the magnitude of this benefit depended on cue implementation. Haptic and auditory guidance substantially reduced post-acquisition elapsed time in both SPV conditions, and these effects remained significant in the stricter sensitivity analysis. The corresponding visual-guidance effect was less robust. 
Thus, the whole-trial benefit cannot be assigned primarily to acquisition: haptic and auditory guidance improved both initial orienting and subsequent search completion.

Accumulated in-view dwell provides a complementary view of target maintenance rather than a second stage duration.
Because it excludes periods in which the target center left the horizontal \ac{FOV}, longer dwell can reflect better maintenance of alignment rather than slower processing. 
Consistent with this interpretation, guidance reduced post-acquisition out-of-view time, including substantial reductions under visual guidance. 
The acquisition/post-acquisition decomposition therefore provides the cleaner stage analysis, while dwell and re-exits characterize how well alignment was maintained after first entry.

The stage split also separates directional guidance from scene-processing approaches in prosthetic vision.
Image simplification, semantic segmentation, and depth filtering improve the information represented through the implant
~\cite{vergnieux_simplification_2017,
sanchez-garcia_semantic_2020,rasla_relative_2022,han_deep_2021,kasowskiStaticTemporalSemantic2025}.
Directional guidance reduces the search required to bring relevant
content into that representation.
Our data separate these two costs within an active search task.

\subsection{Cross-Modal Guidance Under Severe Visual Constraints}

Visual guidance is often highly effective for out-of-view search.
In limited-FOV \ac{AR}, Marquardt et al.~
\cite{marquardt_comparing_2020} found faster search with visual
EyeSee360 than with audio-tactile guidance, although audio-tactile
guidance better preserved situation awareness.
Other work likewise shows that cue effectiveness depends on task demands
and available sensory evidence
~\cite{jeong_differential_2024,sawahata_synergy_2024}.

Our cue implementations showed a different ordering under SPV.
All three modalities conveyed the same horizontal target offset and
reduced excess yaw to a similar degree
({\bf Figure~\ref{fig:results}F}), yet auditory and haptic guidance
produced 11--13\% faster acquisition and approximately 25\% faster
overall search than visual guidance
({\bf Figure~\ref{fig:results}A,B}).
These differences were similar across the $10\times10$ and $20\times20$ conditions, indicating that the nonvisual advantage generalized across both SPV regimes rather than increasing with coarser vision.
The nonvisual advantage therefore reflected faster use of the guidance
signal rather than a more direct orienting trajectory.
This is consistent with work showing benefits for guidance that can be
translated directly into movement decisions
~\cite{varshney_actionable_2026}.

The post-acquisition results do not support a simple visual-interference
account.
At $20\times20$, visual guidance did not reliably reduce
post-acquisition time, whereas haptic and auditory guidance did; at
$10\times10$, all three cues shortened this stage.
Participant reports nevertheless suggest that the visual cue could
compete with scene content, consistent with prior work on visual clutter
and occlusion~\cite{gruenefeld_improving_2019,marquardt_comparing_2020,
kumaran_impact_2023,kim_go_2025,kudnick_ripplevision_2026}.

These results should not be interpreted as an intrinsic ranking of
sensory modalities.
The cues encoded the same target-offset variable using modality-specific mappings rather than physically matched parameters, since identical temporal or intensity mappings would not imply perceptual equivalence across vision, audition, and touch.
Differences in salience, discriminability, or sensorimotor mapping
could therefore contribute to the observed ordering.
The results establish the relative performance of the tested
implementations, not an intrinsic advantage of one sensory modality
over another.

\subsection{Design Implications for Constrained XR}

Cross-modal guidance may be especially useful when the visual channel already carries task-relevant content.
Visual guidance can work well when it does not compete with that content
~\cite{marquardt_comparing_2020,kudnick_ripplevision_2026}, whereas in both SPV conditions tested here, auditory and haptic guidance provided effective orienting without occupying the visual representation.
Whether this advantage grows as visual bandwidth decreases remains an open question.

Guidance may also benefit from adapting across search stages.
All cues improved initial acquisition, while post-acquisition benefits were strongest for haptic and auditory guidance.
A system could therefore provide strong directional information while the target is out of view, then reduce or change the cue after acquisition.

Finally, coarse guidance may be sufficient.
The cues conveyed only horizontal target offset, yet removed most excess pre-acquisition rotation
({\bf Figure~\ref{fig:results}F}).
At $10\times10$, guidance also reduced final head-target offset by 52--61\% and improved elevation error despite providing no vertical information.
For constrained displays, the most useful augmentation may therefore be the smallest signal needed to resolve the current spatial uncertainty.

\subsection{Limitations and Future Work}

Our study used sighted participants experiencing \ac{SPV} over a short experimental session.
\ac{SPV} enables controlled manipulation of guidance, clutter, and target geometry while preserving active head-directed search, but it does not reproduce blindness, long-term adaptation, or the idiosyncratic percepts of individual implants.
Validation with visual-prosthesis users will therefore be important for determining how well these findings generalize to clinical use.

Our cross-modal comparison held the underlying target-direction information constant, but the cue implementations necessarily differed in their perceptual and temporal mappings.
The resulting performance ordering should therefore be interpreted as a comparison of the tested implementations rather than an intrinsic ranking of sensory modalities.
Future work should test alternative mappings, including guidance that adapts after target acquisition rather than remaining fixed throughout the trial.

The task also assumed perfect knowledge of a stationary target in a seated workspace and provided only horizontal guidance.
Real systems must contend with greater spatial complexity, uncertain detections, occlusion, moving or absent targets, and locomotion.
Future systems could use scene information to support initial acquisition and then adapt guidance as the user transitions to target localization and identification. 
This coordination between directional guidance and scene-processing methods should be tested under more realistic conditions and, ultimately, with real visual-prosthesis users.







\clearpage
\section*{Acknowledgments}
Supported by the National Library of Medicine of the National Institutes of
Health (NIH) under Award Number DP2-LM014268. The content is solely the
responsibility of the authors and does not necessarily represent the official views
of the NIH.

AI disclosure: OpenAI ChatGPT (GPT-5.6 Sol) was used to assist with manuscript revision, LaTeX table generation, and development of analysis code.
Anthropic Claude Opus 5 was used to critically review the Methods and Results for potential methodological, statistical, and reporting issues.
The authors reviewed all AI-assisted material, verified the analyses, and made the final decisions about the manuscript and its interpretation.

\bibliographystyle{abbrv-doi}

\bibliography{references}

\end{document}